\documentclass[preprint,pre,showpacs,showkeys]{revtex4}%
\usepackage{amssymb}
\usepackage{amsmath}
\usepackage{amsfonts}
\usepackage{graphicx}%
\newtheorem{theorem}{Theorem}
\newtheorem{acknowledgement}[theorem]{Acknowledgement}

\begin{document}
\title[ ]{$q$-Stirling's formula in Tsallis statistics}
\author{Hiroki Suyari}
\email{suyari@ieee.org, suyari@faculty.chiba-u.jp}
\affiliation{Department of Information and Image Sciences, Faculty of Engineering, Chiba
University, 263-8522, Japan}

\begin{abstract}
We present the $q$-Stirling's formula using the $q$-product determined by
Tsallis entropy as nonextensive generalization of the usual Stirling's
formula. The numerical computations and the proof are also shown. Moreover, we
derive and prove some inequalities and equalities concerning the
$q$-Stirling's formula. Applying the $q$-Stirling's formula, the mathematical
foundations of Tsallis statistics can be obtained as similarly as those of
Boltzmann-Gibbs statistics.

\end{abstract}
\eid{ }
\date{\today }
\startpage{1}
\endpage{1}
\maketitle

\section{Introduction}

Tsallis statistics derived from Tsallis entropy \cite{Ts88}\cite{CT91}
provides many successful descriptions of the physical systems with power-law
behavior, which cannot be systematically recovered in the usual
Boltzmann-Gibbs statistical mechanics \cite{AO01}. The appearance of such a
new physics stimulates us to finding its corresponding new mathematical
structure behind it \cite{Su02a}. On the way to the goal, we obtain the
$q$-\textit{Stirling's formula} as nonextensive generalization of the usual
Stirling's formula.

The $q$-Stirling's formula is derived from the $q$-\textit{product} determined
by Tsallis entropy. The $q$-product is first introduced by Borges in
\cite{Bo03} for satisfying the following equations:%
\begin{align}
\ln_{q}\left(  x\otimes_{q}y\right)   &  =\ln_{q}x+\ln_{q}%
y,\label{property of ln_q}\\
\exp_{q}\left(  x\right)  \otimes_{q}\exp_{q}\left(  y\right)   &  =\exp
_{q}\left(  x+y\right)  ,\label{property of exp_q}%
\end{align}
where $\ln_{q}x$ and $\exp_{q}\left(  x\right)  $ are the $q$%
-\textit{logarithm function\cite{AO01}}:%
\begin{equation}
\ln_{q}x:=\frac{x^{1-q}-1}{1-q}\quad \left(  x>0,q\in \mathbb{R}^{+}\right)
\label{q-logarithm}%
\end{equation}
and its inverse function, the $q$-\textit{exponential function\cite{AO01}}:
\begin{equation}
\exp_{q}\left(  x\right)  :=\left \{
\begin{array}
[c]{ll}%
\left[  1+\left(  1-q\right)  x\right]  ^{\frac{1}{1-q}} & \text{if }1+\left(
1-q\right)  x>0,\\
0 & \text{otherwise}%
\end{array}
\right.  \quad \left(  x\in \mathbb{R},q\in \mathbb{R}^{+}\right)
.\label{q-exponential}%
\end{equation}
\textit{\ }The requirements (\ref{property of ln_q}) and
(\ref{property of exp_q}) lead us to the definition of $\otimes_{q}$ between
two positive numbers%
\begin{equation}
x\otimes_{q}y:=\left \{
\begin{array}
[c]{ll}%
\left[  x^{1-q}+y^{1-q}-1\right]  ^{\frac{1}{1-q}}, & \text{if }%
x>0,\,y>0,\,x^{1-q}+y^{1-q}-1>0\\
0, & \text{otherwise}%
\end{array}
\right.  ,\label{def of q-product}%
\end{equation}
which is called the $q$-\textit{product} in \cite{Bo03}. The $q$-product
recovers the usual product:
\begin{equation}
\underset{q\rightarrow1}{\lim}\left(  x\otimes_{q}y\right)  =xy.
\end{equation}
The fundamental properties of the $q$-product $\otimes_{q}$ are almost the
same as the usual product, but the distributive law does not hold in general.%
\begin{equation}
a\left(  x\otimes_{q}y\right)  \neq ax\otimes_{q}y\quad \left(  a\in
\mathbb{R}^{+},x,y\in \mathbb{R}^{+}\right)
\end{equation}
In our previous paper \cite{ST04a}, a \textit{Tsallis distribution} is derived
from applying the $q$-product to the likelihood function in \textit{maximum
likelihood principle} (MLP for short) instead of the usual product. This first
successful application of the $q$-product to the MLP is briefly reviewed as follows.

Consider the same setting as Gauss' law of error: we get $n$ observed values:%
\begin{equation}
x_{1},x_{2},\cdots,x_{n}\in \mathbb{R}%
\end{equation}
as results of $n$ measurements for some observation. Each observed value
$x_{i}\, \left(  i=1,\cdots,n\right)  $ is each value of identically
distributed random variable $X_{i}\, \left(  i=1,\cdots,n\right)  $,
respectively. There exist a true value $x$ satisfying the \textit{additive}
relation:%
\begin{equation}
x_{i}=x+e_{i}\quad \left(  i=1,\cdots,n\right)  ,
\end{equation}
where each of $e_{i}$ is an error in each observation of a true value $x$.
Thus, for each $X_{i}$, there exists a random variable $E_{i}$ such that
$X_{i}=x+E_{i}\, \left(  i=1,\cdots,n\right)  $. Every $E_{i}$ has the same
probability density function $f$ which is differentiable, because
$X_{1},\cdots,X_{n}$ are identically distributed random variables. (i.e.,
$E_{1},\cdots,E_{n}$ are also identically distributed random variables.) Let
$L_{q}\left(  \theta \right)  $ be a function of a variable $\theta,$ defined
by%
\begin{equation}
L_{q}\left(  \theta \right)  =L_{q}\left(  x_{1},x_{2},\cdots,x_{n}%
;\theta \right)  :=f\left(  x_{1}-\theta \right)  \otimes_{q}f\left(
x_{2}-\theta \right)  \otimes_{q}\cdots \otimes_{q}f\left(  x_{n}-\theta \right)
.
\end{equation}
If the function $L_{q}\left(  x_{1},x_{2},\cdots,x_{n};\theta \right)  $ of
$\theta$ for any fixed $x_{1},x_{2},\cdots,x_{n}$ takes the maximum at%
\begin{equation}
\theta=\theta^{\ast}:=\frac{x_{1}+x_{2}+\cdots+x_{n}}{n},
\end{equation}
then the probability density function $f$ must be a \textit{Tsallis
distribution}:%
\begin{equation}
f\left(  e\right)  =\frac{\exp_{q}\left(  -\beta_{q}e^{2}\right)  }{\int
\exp_{q}\left(  -\beta_{q}e^{2}\right)  de}%
\end{equation}
where $\beta_{q}$ is a $q$-dependent positive constant. This result recovers
Gauss' law of error when $q\rightarrow1$. See \cite{ST04a} for the proof.

In addition to the above successful application of the $q$-product, we show
one more important property of the $q$-product in section II. This property
convinces us of the undoubted validity of the $q$-product in Tsallis
statistics. Along the lines of the above successful application and the
important properties of the $q$-product, we derive the $q$-Stirling's formula
with the proof and numerical computations in section III.

\section{Representation of the $q$-exponential function by means of the
$q$-product}

The base of the natural logarithm $e$ is well-known to be defined as%
\begin{equation}
\exp \left(  1\right)  :=\underset{n\rightarrow \infty}{\lim}\left(  1+\frac
{1}{n}\right)  ^{n}.\label{def of e}%
\end{equation}
Using this definition, the exponential function $\exp \left(  x\right)  $ is
expressed by%
\begin{equation}
\exp \left(  x\right)  =\underset{n\rightarrow \infty}{\lim}\left(  1+\frac
{x}{n}\right)  ^{n}.\label{def of expx}%
\end{equation}
Replacing the power on the right side of (\ref{def of expx}) by the $n$ times
of the $q$-product $\otimes_{q}^{n}:$%
\begin{equation}
x^{\otimes_{q}^{n}}:=\underset{n\text{ times}}{\underbrace{x\otimes_{q}%
\cdots \otimes_{q}x}},
\end{equation}
$\exp_{q}\left(  x\right)  $ is obtained. In other words, $\underset
{n\rightarrow \infty}{\lim}\left(  1+\frac{x}{n}\right)  ^{\otimes_{q}^{n}}$
coincides with $\exp_{q}\left(  x\right)  .$
\begin{equation}
\exp_{q}\left(  x\right)  =\underset{n\rightarrow \infty}{\lim}\left(
1+\frac{x}{n}\right)  ^{\otimes_{q}^{n}}\label{repre of q-exp}%
\end{equation}
$\exp_{q}\left(  x\right)  $ in (\ref{q-exponential}) was originally
introduced in \cite{Ts94} from Tsallis entropy and its maximization. On the
other hand, $\underset{n\rightarrow \infty}{\lim}\left(  1+\frac{x}{n}\right)
^{\otimes_{q}^{n}}$ is computed using the definition of the $q$-product only.
Therefore, this coincidence is an evidence revealing the conclusive validity
of the $q$-product in Tsallis statistics as similarly as the law of error in
Tsallis statistics. Therefore, the $q$-Stirling's formula using the
$q$-product is naturally defined and computed in the following section. The
proof of (\ref{repre of q-exp}) is given in the appendix.

\section{$q$-Stirling's formula}

\subsection{$q$-factorial $n!_{q}$}

\bigskip The $q$-factorial $n!_{q}\,$\ for $n\in \mathbb{N}$ and $q>0$ is
defined by%
\begin{equation}
n!_{q}:=1\otimes_{q}\cdots \otimes_{q}n\label{def of q-kaijyo}%
\end{equation}
where $\otimes_{q}$ is the $q$-product (\ref{def of q-product}). Using the
definition (\ref{def of q-product}), $n!_{q}\,$\ is explicitly expressed by%
\begin{equation}
n!_{q}=\left[  \sum_{k=1}^{n}k^{1-q}-\left(  n-1\right)  \right]  ^{\frac
{1}{1-q}}.\label{explicit form of q-kaijyo}%
\end{equation}
This implies%
\begin{equation}
\ln_{q}\left(  n!_{q}\right)  =\frac{\sum \limits_{k=1}^{n}k^{1-q}-n}%
{1-q}.\label{explicit form of lnqn!}%
\end{equation}
In the present paper, we derive the approximation of $\ln_{q}\left(
n!_{q}\right)  $ instead of that of $n!_{q}$. In fact, when $q>1$, there
exists infinite numbers of $n\in \mathbb{N}$ such that $\sum_{k=1}^{n}%
k^{1-q}-\left(  n-1\right)  <0$. For example,
\begin{equation}
\sum_{k=1}^{n}k^{1-q}-\left(  n-1\right)  <0\quad \text{when}\quad
q=1.2\quad \text{and}\quad n\geq6.
\end{equation}
When $\sum_{k=1}^{n}k^{1-q}-\left(  n-1\right)  <0$, $n!_{q}$ takes complex
numbers in general except $q=1+\frac{1}{2m}\left(  m\in \mathbb{N}\right)  $.
On the other hand, when $\sum_{k=1}^{n}k^{1-q}-\left(  n-1\right)  >0$ for a
given $n\in \mathbb{N}$ and $q>0$, we easily obtain $n!_{q}$ from $\ln
_{q}\left(  n!_{q}\right)  $. Therefore, the approximation of $\ln_{q}\left(
n!_{q}\right)  $ is called \textquotedblleft$q$-Stirling's
formula\textquotedblright \ in the present paper. If an approximation of
$\ln_{q}\left(  n!_{q}\right)  $ is not needed, the above explicit form
(\ref{explicit form of lnqn!}) should be directly used for its computation.
However, in order to clarify the correspondence between the studies $q=1$ and
$q\neq1$, the approximation of $\ln_{q}\left(  n!_{q}\right)  $ is useful. In
fact, using the present $q$-Stirling's formula, we obtain the surprising
mathematical structure in Tsallis statistics \cite{ST04b}.

We obtain the $q$-Stirling's formula as follows:
\begin{equation}
\ln_{q}\left(  n!_{q}\right)  =\left \{
\begin{array}
[c]{ll}%
\left(  n+\frac{1}{2}\right)  \ln n+\left(  -n\right)  +\theta_{n,1}+\left(
1-\delta_{1}\right)  \quad & \text{if}\quad q=1,\\
n-\frac{1}{2n}-\ln n-\frac{1}{2}+\theta_{n,2}-\delta_{2} & \text{if}\quad
q=2,\\
\left(  \frac{n}{2-q}+\frac{1}{2}\right)  \frac{n^{1-q}-1}{1-q}+\left(
-\frac{n}{2-q}\right)  +\theta_{n,q}+\left(  \frac{1}{2-q}-\delta_{q}\right)
& \text{if}\quad q>0\quad \text{and}\quad q\neq1,2,
\end{array}
\right. \label{q-Stirling formula}%
\end{equation}
where%
\begin{equation}
\underset{n\rightarrow \infty}{\lim}\theta_{n,q}=0,\quad0<\theta_{n,1}<\frac
{1}{12n},\quad e^{1-\delta_{1}}=\sqrt{2\pi}.\label{properties of parameters}%
\end{equation}
It is easily verified that the $q$-Stirling's formula coincides with the usual
Stirling's formula when $q\rightarrow1$.%
\begin{equation}
\underset{q\rightarrow1}{\lim}\ln_{q}\left(  n!_{q}\right)  =\ln n!
\end{equation}

Other properties of $\theta_{n,q}$ and $\delta_{q}$ such as explicit forms and
approximations are still unclear. But numerical results of
(\ref{q-Stirling formula}) exhibit good approximations of the true value
(\ref{explicit form of lnqn!}), which is shown in the next subsection.

\subsection{Numerical computation of the $q$-Stirling's formula}

\bigskip By the property $\underset{n\rightarrow \infty}{\lim}\theta_{n,q}=0$
in (\ref{properties of parameters}), $\theta_{n,q}$ is set to be $0$ for the
computation. The explicit form of $\delta_{q}$ is not clear yet, but
$\delta_{q}$ does \textit{not} depend on $n$. Thus, $\delta_{q}$ is set to be
$\delta_{q}=\delta_{1}=1-\ln \sqrt{2\pi}\simeq0.081$ which is already known in
the usual Stirling's formula. Then the $q$-Stirling's formula for the
numerical computations in this subsection is given by%
\begin{equation}
\ln_{q}\left(  n!_{q}\right)  \simeq \left \{
\begin{array}
[c]{ll}%
\left(  n+\frac{1}{2}\right)  \ln n+\left(  -n\right)  +\left(
1-0.081\right)  \quad & \text{if}\quad q=1,\\
n-\frac{1}{2n}-\ln n-\frac{1}{2}-0.081 & \text{if}\quad q=2,\\
\left(  \frac{n}{2-q}+\frac{1}{2}\right)  \frac{n^{1-q}-1}{1-q}+\left(
-\frac{n}{2-q}\right)  +\left(  \frac{1}{2-q}-0.081\right)  & \text{if}\quad
q>0\quad \text{and}\quad q\neq1,2,
\end{array}
\right. \label{suuchikeisann0}%
\end{equation}
According to the following numerical results, these assumptions on
$\theta_{n,q}=0$ and $\delta_{q}=0.081$ are found to be justified when
$n\rightarrow \infty$. We compute the true value (\ref{explicit form of lnqn!})
and its approximate value (\ref{suuchikeisann0}) for $\ln_{q}\left(
n!_{q}\right)  $. Moreover, the relative error and the absolute error are also
computed.%
\begin{align}
\text{relative error}  & :=\frac{\left \vert \text{true value}%
-\text{approximate value}\right \vert }{\text{true value}},\label{relative}\\
\text{absolute error}  & :=\left \vert \text{true value}-\text{approximate
value}\right \vert .\label{absolute}%
\end{align}%
\begin{equation}
q=0.1\, \, \, \, \,%
\begin{tabular}
[c]{c|c|c|c|c}\hline
$n$ & $\ln_{q}\left(  n!_{q}\right)  $ $\text{(}%
\ref{explicit form of q-kaijyo}\text{)}$ & approximate value
$\text{(\ref{suuchikeisann0})}$ & relative error & absolute error\\ \hline
$2$ & $9.6230\times10^{-1}$ & $8.8677\times10^{-1}$ & $7.8480\times10^{-2} $ &
$7.5521\times10^{-2}$\\
$5$ & $9.2143$ & $9.1455$ & $7.4604\times10^{-3}$ & $6.8742\times10^{-2}$\\
$10$ & $3.9708\times10^{1}$ & $3.9644\times10^{1}$ & $1.6117\times10^{-3}$ &
$6.3995\times10^{-2}$\\
$15$ & $9.0005\times10^{1}$ & $8.9943\times10^{1}$ & $6.8181\times10^{-4}$ &
$6.1366\times10^{-2}$\\
$20$ & $1.5933\times10^{2}$ & $1.5927\times10^{2}$ & $3.7384\times10^{-4}$ &
$5.9563\times10^{-2}$\\
$\downarrow$ &  &  & $\downarrow$ & $\downarrow$\\
$\infty$ &  &  & $0$ & $0$\\ \hline
\end{tabular}
\label{q=0.1}%
\end{equation}%
\begin{equation}
q=0.5\, \, \, \, \,%
\begin{tabular}
[c]{c|c|c|c|c}\hline
$n$ & $\ln_{q}\left(  n!_{q}\right)  $ $\text{(}%
\ref{explicit form of q-kaijyo}\text{)}$ & approximate value
$\text{(\ref{suuchikeisann0})}$ & relative error & absolute error\\ \hline
$2$ & $8.2843\times10^{-1}$ & $7.7112\times10^{-1}$ & $6.9180\times10^{-2} $ &
$5.7311\times10^{-2}$\\
$5$ & $6.7647$ & $6.7289$ & $5.2937\times10^{-3}$ & $3.5810\times10^{-2}$\\
$10$ & $2.4937\times10^{1}$ & $2.4912\times10^{1}$ & $9.9893\times10^{-4}$ &
$2.4910\times10^{-2}$\\
$15$ & $5.0938\times10^{1}$ & $5.0918\times10^{1}$ & $3.9413\times10^{-4}$ &
$2.0076\times10^{-2}$\\
$20$ & $8.3332\times10^{1}$ & $8.3315\times10^{1}$ & $2.0633\times10^{-4}$ &
$1.7194\times10^{-2}$\\
$\downarrow$ &  &  & $\downarrow$ & $\downarrow$\\
$\infty$ &  &  & $0$ & $0$\\ \hline
\end{tabular}
\end{equation}%
\begin{equation}
q=0.999\, \, \, \, \,%
\begin{tabular}
[c]{c|c|c|c|c}\hline
$n$ & $\ln_{q}\left(  n!_{q}\right)  $ $\text{(}%
\ref{explicit form of q-kaijyo}\text{)}$ & approximate value
$\text{(\ref{suuchikeisann0})}$ & relative error & absolute error\\ \hline
$2$ & $6.9339\times10^{-1}$ & $6.5208\times10^{-1}$ & $5.9570\times10^{-2} $ &
$4.1305\times10^{-2}$\\
$5$ & $4.7906$ & $4.7740$ & $3.4666\times10^{-3}$ & $1.6607\times10^{-2}$\\
$10$ & $1.5118\times10^{1}$ & $1.5110\times10^{1}$ & $5.4802\times10^{-4}$ &
$8.2851\times10^{-3}$\\
$15$ & $2.7930\times10^{1}$ & $2.7924\times10^{1}$ & $1.9711\times10^{-4}$ &
$5.5052\times10^{-3}$\\
$20$ & $4.2387\times10^{1}$ & $4.2383\times10^{1}$ & $9.7063\times10^{-5}$ &
$4.1142\times10^{-3}$\\
$\downarrow$ &  &  & $\downarrow$ & $\downarrow$\\
$\infty$ &  &  & 0 & 0\\ \hline
\end{tabular}
\end{equation}%
\begin{equation}
q=1.0\, \, \, \, \,%
\begin{tabular}
[c]{c|c|c|c|c}\hline
$n$ & $\ln \left(  n!\right)  $ & approximate value
$\text{(\ref{suuchikeisann0})}$ & relative error & absolute error\\ \hline
$2$ & $6.9315\times10^{-1}$ & $6.5187\times10^{-1}$ & $5.9553\times10^{-2} $ &
$4.1279\times10^{-2}$\\
$5$ & $4.7875$ & $4.7709$ & $3.4639\times10^{-3}$ & $1.6583\times10^{-2}$\\
$10$ & $1.5104\times10^{1}$ & $1.5096\times10^{1}$ & $5.4746\times10^{-4}$ &
$8.2691\times10^{-3}$\\
$15$ & $2.7899\times10^{1}$ & $2.7894\times10^{1}$ & $1.9690\times10^{-4}$ &
$5.4933\times10^{-3}$\\
$20$ & $4.2336\times10^{1}$ & $4.2332\times10^{1}$ & $9.6960\times10^{-5}$ &
$4.1049\times10^{-3}$\\
$\downarrow$ &  &  & $\downarrow$ & $\downarrow$\\
$\infty$ &  &  & $0$ & $0$\\ \hline
\end{tabular}
\end{equation}%
\begin{equation}
q=1.5\, \, \, \, \,%
\begin{tabular}
[c]{c|c|c|c|c}\hline
$n$ & $\ln_{q}\left(  n!_{q}\right)  $ $\text{(}%
\ref{explicit form of q-kaijyo}\text{)}$ & approximate value
$\text{(\ref{suuchikeisann0})}$ & relative error & absolute error\\ \hline
$2$ & $5.8579\times10^{-1}$ & $5.5504\times10^{-1}$ & $5.2489\times10^{-2} $ &
$3.0747\times10^{-2}$\\
$5$ & $3.5367$ & $3.5275$ & $2.5856\times10^{-3}$ & $9.1442\times10^{-3}$\\
$10$ & $9.9580$ & $9.9537$ & $4.3609\times10^{-4}$ & $4.3426\times10^{-3}$\\
$15$ & $1.7172\times10^{1}$ & $1.7169\times10^{1}$ & $1.8303\times10^{-4}$ &
$3.1431\times10^{-3}$\\
$20$ & $2.4809\times10^{1}$ & $2.4807\times10^{1}$ & $1.0643\times10^{-4}$ &
$2.6406\times10^{-3}$\\
$\downarrow$ &  &  & $\downarrow$ & $\downarrow$\\
$\infty$ &  &  & $0$ & $0$\\ \hline
\end{tabular}
\end{equation}%
\begin{equation}
q=2.0\, \, \, \, \,%
\begin{tabular}
[c]{c|c|c|c|c}\hline
$n$ & $\ln_{q}\left(  n!_{q}\right)  $ $\text{(}%
\ref{explicit form of q-kaijyo}\text{)}$ & approximate value
$\text{(\ref{suuchikeisann0})}$ & relative error & absolute error\\ \hline
$2$ & $5.0000\times10^{-1}$ & $4.7585\times10^{-1}$ & $4.8294\times10^{-2} $ &
$2.4147\times10^{-2}$\\
$5$ & $2.7167$ & $2.7096$ & $2.6152\times10^{-3}$ & $7.1046\times10^{-3}$\\
$10$ & $7.0710$ & $7.0664$ & $6.5292\times10^{-4}$ & $4.6168\times10^{-3}$\\
$15$ & $1.1682\times10^{1}$ & $1.1678\times10^{1}$ & $3.5564\times10^{-4}$ &
$4.1545\times10^{-3}$\\
$20$ & $1.6402\times10^{1}$ & $1.6398\times10^{1}$ & $2.4342\times10^{-4}$ &
$3.9926\times10^{-3}$\\
$\downarrow$ &  &  & $\downarrow$ & $\downarrow$\\
$\infty$ &  &  & $0$ & $0$\\ \hline
\end{tabular}
\label{q=2.0}%
\end{equation}%
\begin{figure}
[ptbh]
\begin{center}
\includegraphics[
height=3.1367in,
width=4.2229in
]%
{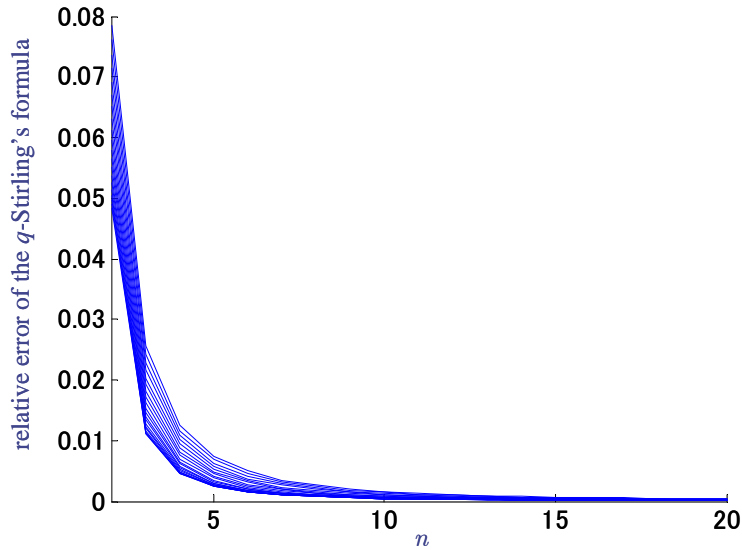}%
\caption{relative error of the $q-$Stirling formula with respect to $n$ when
$q=0.1,0.2,\cdots,2.0$}%
\label{fig: relative error}%
\end{center}
\end{figure}
The numerical results (\ref{q=0.1})$\sim$(\ref{q=2.0}) and
Fig.\ref{fig: relative error} show that the relative error of the
$q$-Stirling's formula (\ref{suuchikeisann0}) in any $q>0$ goes to zero when
$n\rightarrow \infty$. Therefore, the $q$-Stirling's formula
(\ref{q-Stirling formula}) can provide a good approximate value for each
$\ln_{q}\left(  n!_{q}\right)  $ in any $q>0$.

\subsection{\bigskip Derivation of the $q$-Stirling's formula}

This subsection proves the $q$-Stirling's formula (\ref{q-Stirling formula})
when $q>0\, \left(  q\neq1\right)  $. The case $q=1$ of the $q$-Stirling's
formula (\ref{q-Stirling formula}) is the famous usual Stirling's formula
whose proof is found in many references such as \cite{Fe68}, so that we prove
the $q$-Stirling's formula in case $q>0\, \left(  q\neq1\right)  $. For
simplicity, we assume $q\neq1$ in the subsections C and D.

Note that, when $q=0,$ $n!_{q}$ is directly and explicitly derived from
(\ref{explicit form of q-kaijyo}) as follows:
\begin{equation}
n!_{q=0}=\sum_{k=1}^{n}k-\left(  n-1\right)  =\frac{1}{2}n^{2}-\frac{1}{2}n+1.
\end{equation}
Thus, the $q$-Stirling's formula when $q=0$ is not needed. When $q<0,\ln_{q}x$
is a convex function with respect to $x$, so that $\delta_{n,q}$ defined by
(\ref{delta_nq}) below does not converge when $n\rightarrow \infty$ in general.
Moreover, Tsallis entropy when $q<0$ lacks the property of concavity which
plays essential roles in the natural generalization of Boltzmann-Gibbs
statistical mechanics. Therefore, we prove the $q$-Stirling's formula
(\ref{q-Stirling formula}) when $q>0$.

\begin{figure}
[ptbh]
\begin{center}
\includegraphics[
height=3.672in,
width=4.8862in
]%
{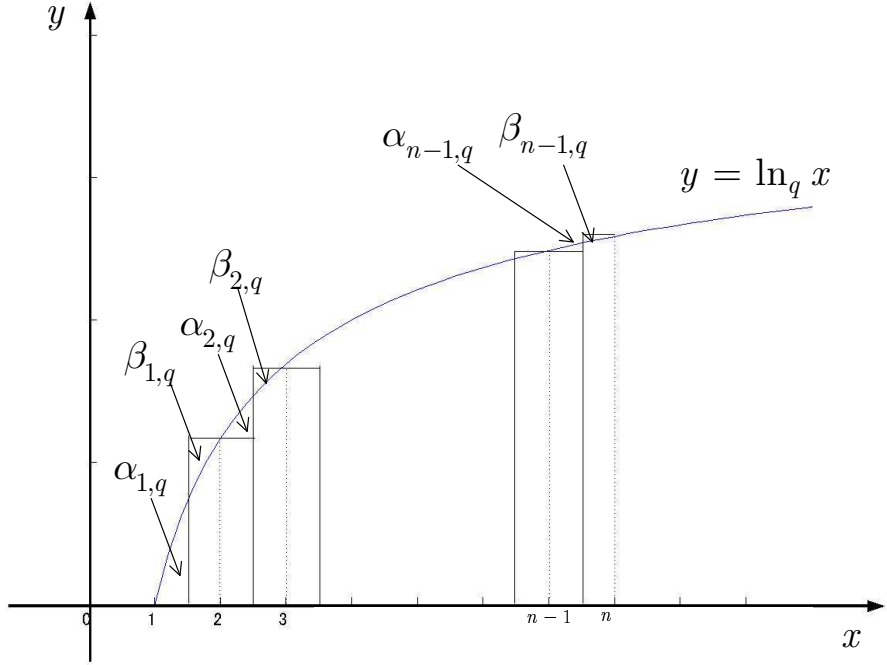}%
\caption{approximation of $\int_{1}^{n}\ln_{q}xdx$}%
\label{fig: approximate}%
\end{center}
\end{figure}

\begin{figure}
[ptbhptbh]
\begin{center}
\includegraphics[
height=2.5417in,
width=3.6729in
]%
{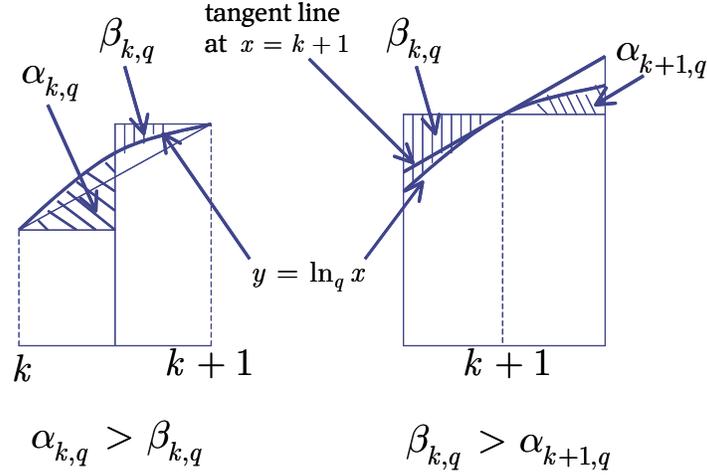}%
\caption{magnitude relation between $\alpha_{k,q}$ and $\beta_{k,q}$}%
\label{fig:mag relation}%
\end{center}
\end{figure}

Consider the definite integral $\int_{1}^{n}\ln_{q}xdx$ when $q>0\, \left(
q\neq1\right)  .$ The curve in Fig.\ref{fig: approximate} describes a concave
function with respect to $x$:%
\begin{equation}
y=\ln_{q}x=\frac{x^{1-q}-1}{1-q}.
\end{equation}
For each natural number $k\in \left \{  2,3,\cdots \right \}  $ on the $x$-axis,
there exist a rectangle with width 1 and height $\ln_{q}k$ (See
Fig.\ref{fig: approximate}). Here, in order to avoid confusions, we call a
rectangle at $k\in \left \{  2,3,\cdots \right \}  $ on the $x$-axis by
\textquotedblleft$k$-rectangle\textquotedblright. The center of the bottom of
each $k$-rectangle is at $k\in \left \{  2,3,\cdots \right \}  $ on the $x$-axis.
For each $k$-rectangle, we define $\alpha_{k,q}$ and $\beta_{k,q}$ for
$k\in \mathbb{N}$ by%
\begin{align}
\alpha_{k,q} &  :=\text{the area of the portion surrounded by the curve, the
top side of the }k\text{-rectangle}\nonumber \\
&  \qquad \text{ (}x\text{-axis when }k=1\text{) and the left side of the
}\left(  k+1\right)  \text{-rectangle,}\\
\beta_{k,q} &  :=\text{the area of the portion surrounded by the curve, the
left side}\nonumber \\
&  \qquad \text{ of the }\left(  k+1\right)  \text{-rectangle and the top side
of the }\left(  k+1\right)  \text{-rectangle.}%
\end{align}%
Then, we define the sum of areas of all gaps between the curve and the
rectangles by $\delta_{n,q}$%
\begin{equation}
\delta_{n,q}:=\sum \limits_{k=1}^{n-1}\alpha_{k,q}-\sum \limits_{k=1}^{n-1}%
\beta_{k,q}=\sum \limits_{k=1}^{n-1}\left(  \alpha_{k,q}-\beta_{k,q}\right)
.\label{delta_nq}%
\end{equation}
Using $\delta_{n,q}$, the area of the portion surrounded by the $x$-axis, the
straight line $x=n$ and the curve $y=\ln_{q}x$ is computed as%
\begin{align}
&  \ln_{q}2+\ln_{q}3+\cdots+\ln_{q}\left(  n-1\right)  +\frac{1}{2}\ln
_{q}n+\delta_{n,q}\\
&  =\int_{1}^{n}\ln_{q}xdx=\int_{1}^{n}\left(  \frac{x^{1-q}-1}{1-q}\right)
dx=\left \{
\begin{array}
[c]{ll}%
-\ln n+n-1 & \text{if}\quad q=2\\
\frac{1}{2-q}\cdot n\cdot \ln_{q}n-\frac{n-1}{2-q} & \text{if}\quad q\neq2
\end{array}
\right.  .\label{integral_ln_q}%
\end{align}
Therefore, we obtain%
\begin{align}
\ln_{q}\left(  n!_{q}\right)   &  =\ln_{q}\left(  1\otimes_{q}\cdots
\otimes_{q}n\right)  =\ln_{q}2+\ln_{q}3+\cdots+\ln_{q}n\\
&  =\left(  \ln_{q}2+\ln_{q}3+\cdots+\ln_{q}\left(  n-1\right)  +\frac{1}%
{2}\ln_{q}n+\delta_{n,q}\right)  +\frac{1}{2}\ln_{q}n-\delta_{n,q}\\
&  =\left \{
\begin{array}
[c]{ll}%
\left(  -\ln n+n-1\right)  +\frac{1}{2}\left(  1-\frac{1}{n}\right)
-\delta_{n,2} & \text{if}\quad q=2\\
\left(  \frac{1}{2-q}\cdot n\cdot \ln_{q}n-\frac{n-1}{2-q}\right)  +\frac{1}%
{2}\ln_{q}n-\delta_{n,q} & \text{if}\quad q\neq2
\end{array}
\right. \\
&  =\left \{
\begin{array}
[c]{ll}%
n-\frac{1}{2n}-\ln n-\frac{1}{2}-\delta_{n,2} & \text{if}\quad q=2\\
\left(  \frac{n}{2-q}+\frac{1}{2}\right)  \ln_{q}n+\left(  -\frac{n}%
{2-q}\right)  +\left(  \frac{1}{2-q}-\delta_{n,q}\right)  & \text{if}\quad
q\neq2
\end{array}
\right.  .\label{q-n!(1)}%
\end{align}
$\ln_{q}x$ is a concave function of $x$, so that the magnitude relation
between $\alpha_{k,q}$ and $\beta_{k,q}$ is shown in
Fig.\ref{fig:mag relation}.%
From Fig.\ref{fig:mag relation}, we find%
\begin{equation}
\alpha_{1,q}>\beta_{1,q}>\alpha_{2,q}>\cdots>\alpha_{n,q}>\beta_{n,q}%
>\cdots.\label{koutai_series}%
\end{equation}
This means that $``\alpha_{1,q},\beta_{1,q},\alpha_{2,q},\cdots,\alpha
_{n,q},\beta_{n,q},\cdots \textquotedblright$ is a monotone decreasing sequence
with lower bound $0$, so that we construct a sequence of terms of an
alternating series such that%
\begin{equation}
\alpha_{1,q},\,-\beta_{1,q},\, \, \alpha_{2,q},\,-\beta_{2,q},\, \cdots
,\, \alpha_{n,q},\,-\beta_{n,q},\, \cdots.\label{koutai_terms}%
\end{equation}
Thus, $\delta_{n,q}$ defined by (\ref{delta_nq}) is found to be a finite
partial sum of terms of the above sequence (\ref{koutai_terms}). Therefore,
\begin{equation}
\underset{n\rightarrow \infty}{\lim}\delta_{n,q}=\sum \limits_{k=1}^{\infty
}\left(  \alpha_{k,q}-\beta_{k,q}\right)
\end{equation}
is an alternating series. Note that in general an alternating series is of the
form $\sum_{i=1}^{\infty}\left(  -1\right)  ^{i+1}a_{i}\, \, \left(
a_{i}>0\right)  $. It is known that for an alternating series with property
(\ref{koutai_series}) $\delta_{n,q}$ converges as $n\rightarrow \infty$ for any
fixed $q>0$. Thus, we define $\delta_{q}$ and $\theta_{n,q}$ as
\begin{align}
\delta_{q} &  :=\underset{n\rightarrow \infty}{\lim}\delta_{n,q},\\
\theta_{n,q} &  :=\delta_{q}-\delta_{n,q}.
\end{align}
Substituting $\delta_{n,q}=\delta_{q}-\theta_{n,q}$ into (\ref{q-n!(1)})
yields%
\begin{equation}
\ln_{q}\left(  n!_{q}\right)  =\left \{
\begin{array}
[c]{ll}%
n-\frac{1}{2n}-\ln n-\frac{1}{2}-\left(  \delta_{2}-\theta_{n,2}\right)  &
\text{if}\quad q=2\\
\left(  \frac{n}{2-q}+\frac{1}{2}\right)  \ln_{q}n+\left(  -\frac{n}%
{2-q}\right)  +\left(  \frac{1}{2-q}-\left(  \delta_{q}-\theta_{n,q}\right)
\right)  & \text{if}\quad q\neq2
\end{array}
\right.
\end{equation}
where $\delta_{q}=\underset{n\rightarrow \infty}{\lim}\delta_{n,q}$ implies
that%
\begin{equation}
\underset{n\rightarrow \infty}{\lim}\theta_{n,q}=\underset{n\rightarrow \infty
}{\lim}\left(  \delta_{q}-\delta_{n,q}\right)  =0.
\end{equation}
Therefore, when $n$ is sufficiently large, $\ln_{q}\left(  n!_{q}\right)  $ is
approximated as
\begin{equation}
\ln_{q}\left(  n!_{q}\right)  \simeq \left \{
\begin{array}
[c]{ll}%
n-\frac{1}{2n}-\ln n-\frac{1}{2}-\delta_{2} & \text{if}\quad q=2\\
\left(  \frac{n}{2-q}+\frac{1}{2}\right)  \frac{n^{1-q}-1}{1-q}+\left(
-\frac{n}{2-q}\right)  +\left(  \frac{1}{2-q}-\delta_{q}\right)  &
\text{if}\quad q\neq2
\end{array}
\right.  .\label{toriaezu0}%
\end{equation}
The remained problem is to determine $\delta_{q}\,$ which does \textit{not}
depend on $n$. This is still an open problem, but the exact value of
$\delta_{q}$ for actual applications is not needed. The assumption $\delta
_{q}=\delta_{1}=1-\ln \sqrt{2\pi}$ gives sufficiently useful approximations of
the $q$-Stirling's formula (\ref{q-Stirling formula}) as shown in numerical
computations in the previous subsection.

\subsection{Inequalities and equalities concerning the $q$-Stirling's formula}

This subsection presents the following inequalities (\ref{q-inequality2-1}%
)$\sim$(\ref{q-inequality2-2}) and equalities (\ref{q-order1})$\sim
$(\ref{q-order2}) concerning the $q$-Stirling's formula. In some applications,
these formulas are more useful than the $q$-Stirling's formula
(\ref{q-Stirling formula}).
\begin{align}
\frac{n}{2-q}\ln_{q}n-\frac{n}{2-q}+\frac{1}{2}\ln_{q}n+\frac{1}{2-q}-\frac
{1}{8}  & <\ln_{q}\left(  n!_{q}\right) \nonumber \\
\! \!  & <\frac{n}{2-q}\ln_{q}n-\frac{n}{2-q}+\frac{1}{2}\ln_{q}n+\frac{1}%
{2-q}\text{\quad}\left(  \text{if}\quad q\neq2\right)  ,\nonumber \\
& \label{q-inequality2-1}\\
n-\ln n-\frac{1}{2n}-\frac{5}{8}  & <\ln_{q}\left(  n!_{q}\right)  <n-\ln
n-\frac{1}{2n}-\frac{1}{2}\text{\quad}\left(  \text{if}\quad q=2\right)
,\label{q-inequality2-2}%
\end{align}
where $n\in \mathbb{N}$ and $q>0$. These inequalities or the $q$-Stirling's
formula (\ref{q-Stirling formula}) is reduced to the equalities:%
\begin{align}
\ln_{q}\left(  n!_{q}\right)   & =\frac{n}{2-q}\ln_{q}n-\frac{n}{2-q}+O\left(
\ln_{q}n\right)  \text{\quad}\left(  \text{if}\quad q\neq2\right)
,\label{q-order1}\\
\ln_{q}\left(  n!_{q}\right)   & =n-\ln n+O\left(  1\right)  \text{\quad
}\left(  \text{if}\quad q=2\right)  .\label{q-order2}%
\end{align}
These formulas are proved in the following way.
\begin{figure}
[ptbh]
\begin{center}
\includegraphics[
height=2.9879in,
width=4.0171in
]%
{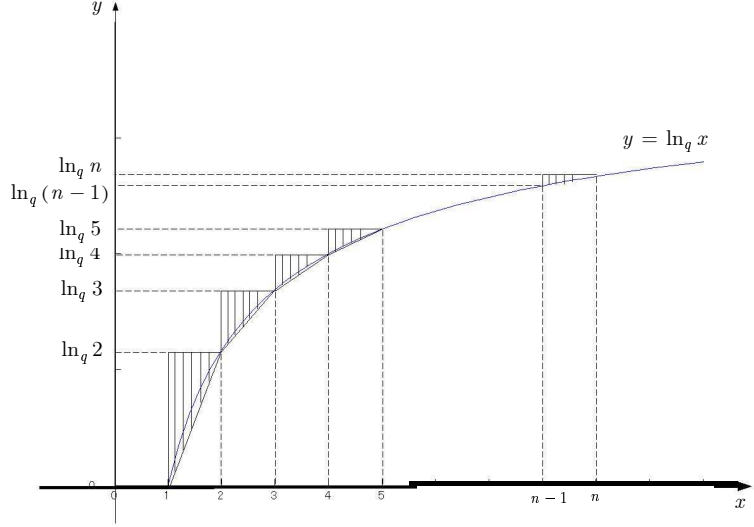}%
\caption{$\int_{1}^{n}\ln_{q}xdx$ and the sum of the triangular areas}%
\label{fig:gray integral}%
\end{center}
\end{figure}

In Fig.\ref{fig:gray integral}, the sum of gray triangular areas is%
\begin{equation}
\frac{1}{2}\ln_{q}2+\frac{1}{2}\left(  \ln_{q}3-\ln_{q}2\right)  +\frac{1}%
{2}\left(  \ln_{q}4-\ln_{q}3\right)  +\cdots+\frac{1}{2}\left(  \ln_{q}%
n-\ln_{q}\left(  n-1\right)  \right)  =\frac{1}{2}\ln_{q}n.
\end{equation}
On the other hand,
\begin{equation}
\ln_{q}\left(  n!_{q}\right)  =\ln_{q}2+\ln_{q}3+\cdots+\ln_{q}n
\end{equation}
implies that $\ln_{q}\left(  n!_{q}\right)  $ represents the sum of the
rectangular areas in Fig.\ref{fig:gray integral}. Thus, from
Fig.\ref{fig:gray integral} we can read that the difference between $\ln
_{q}\left(  n!_{q}\right)  $ and $\int_{1}^{n}\ln_{q}xdx$ is less than the sum
of gray triangular areas.%
\begin{equation}
\ln_{q}\left(  n!_{q}\right)  -\int_{1}^{n}\ln_{q}xdx<\frac{1}{2}\ln_{q}n
\end{equation}
Therefore, we obtain an upper bound of $\ln_{q}\left(  n!_{q}\right)  $ as
follows
\begin{align}
\ln_{q}\left(  n!_{q}\right)   & <\int_{1}^{n}\ln_{q}xdx+\frac{1}{2}\ln_{q}n\\
& =\left \{
\begin{array}
[c]{ll}%
\left(  -\ln n+n-1\right)  +\frac{1}{2}\left(  1-\frac{1}{n}\right)  &
\text{if}\quad q=2\\
\left(  \frac{1}{2-q}\cdot n\cdot \ln_{q}n-\frac{n-1}{2-q}\right)  +\frac{1}%
{2}\ln_{q}n & \text{if}\quad q\neq2
\end{array}
\right. \\
& =\left \{
\begin{array}
[c]{ll}%
n-\ln n-\frac{1}{2n}-\frac{1}{2} & \text{if}\quad q=2\\
\frac{n}{2-q}\ln_{q}n-\frac{n}{2-q}+\frac{1}{2}\ln_{q}n+\frac{1}{2-q} &
\text{if}\quad q\neq2
\end{array}
\right.  ,\label{kekka1}%
\end{align}
where we used (\ref{integral_ln_q}).

We derive an lower bound of $\ln_{q}\left(  n!_{q}\right)  $.
\begin{figure}
[ptbh]
\begin{center}
\includegraphics[
height=3.2854in,
width=4.0767in
]%
{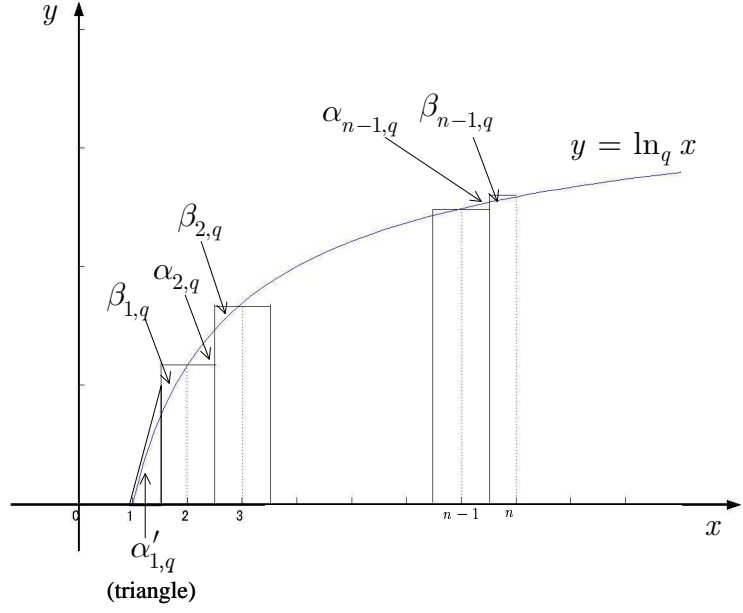}%
\caption{upper bound of $\int_{1}^{n}\ln_{q}xdx $}%
\label{fig: upper bound}%
\end{center}
\end{figure}
In Fig.\ref{fig: upper bound}, the integral $\int_{k-\frac{1}{2}}^{k+\frac
{1}{2}}\ln_{q}xdx\, \left(  k\in \left \{  2,\cdots,n-1\right \}  \right)  $ is
written by means of $\beta_{k-1,q}$ and $\alpha_{k,q}$.%
\begin{equation}
\int_{k-\frac{1}{2}}^{k+\frac{1}{2}}\ln_{q}xdx=\ln_{q}k-\beta_{k-1,q}%
+\alpha_{k,q}\quad \left(  k\in \left \{  2,\cdots,n-1\right \}  \right)
\label{tanzaku0}%
\end{equation}
Using the relation shown in Fig.\ref{fig:mag relation}:%
\begin{equation}
\beta_{k-1,q}>\alpha_{k,q},
\end{equation}
$\alpha_{k,q}$ is replaced by $\beta_{k-1,q}$ in (\ref{tanzaku0}), so that we
get%
\begin{equation}
\int_{k-\frac{1}{2}}^{k+\frac{1}{2}}\ln_{q}xdx<\ln_{q}k-\beta_{k-1,q}%
+\beta_{k-1,q}=\ln_{q}k.\label{tanzaku_ineq0}%
\end{equation}
Summing up the both sides of (\ref{tanzaku_ineq0}) for $k\in \left \{
2,\cdots,n-1\right \}  $,%
\begin{equation}
\int_{\frac{3}{2}}^{n-\frac{1}{2}}\ln_{q}xdx<\ln_{q}2+\cdots+\ln_{q}\left(
n-1\right)  .\label{kukan1}%
\end{equation}
$\alpha_{1,q}^{\prime}$ in Fig.\ref{fig: upper bound} is a triangle surrounded
by the tangent line of $y=\ln_{q}x$ at $x=1$, the $x$-axis and $x=\frac{3}{2}%
$. Thus,%
\begin{equation}
\int_{1}^{\frac{3}{2}}\ln_{q}xdx<\alpha_{1,q}^{\prime}=\frac{1}{8}%
.\label{kukan2}%
\end{equation}
On the other hand,%
\begin{equation}
\int_{n-\frac{1}{2}}^{n}\ln_{q}xdx<\frac{1}{2}\ln_{q}n.\label{kukan3}%
\end{equation}
The sum of each side of (\ref{kukan1}), (\ref{kukan2}) and (\ref{kukan3})
yields%
\begin{align}
\int_{1}^{n}\ln_{q}xdx &  <\frac{1}{8}+\ln_{q}2+\cdots+\ln_{q}\left(
n-1\right)  +\frac{1}{2}\ln_{q}n\\
&  =\ln_{q}\left(  n!_{q}\right)  +\frac{1}{8}-\frac{1}{2}\ln_{q}n.
\end{align}
Thus,%
\begin{equation}
\int_{1}^{n}\ln_{q}xdx+\left(  \frac{1}{2}\ln_{q}n-\frac{1}{8}\right)
<\ln_{q}\left(  n!_{q}\right)  .\label{kibishii0}%
\end{equation}
Substituting (\ref{integral_ln_q}) into $\int_{1}^{n}\ln_{q}xdx$ on the left
side of (\ref{kibishii0}) yields%
\begin{align}
\ln_{q}\left(  n!_{q}\right)   & >\left \{
\begin{array}
[c]{ll}%
\left(  -\ln n+n-1\right)  +\left(  \frac{1}{2}\left(  1-\frac{1}{n}\right)
-\frac{1}{8}\right)  & \text{if}\quad q=2\\
\left(  \frac{1}{2-q}\cdot n\cdot \ln_{q}n-\frac{n-1}{2-q}\right)  +\left(
\frac{1}{2}\ln_{q}n-\frac{1}{8}\right)  & \text{if}\quad q\neq2
\end{array}
\right. \\
& =\left \{
\begin{array}
[c]{ll}%
n-\ln n-\frac{1}{2n}-\frac{5}{8} & \text{if}\quad q=2\\
\frac{n}{2-q}\ln_{q}n-\frac{n}{2-q}+\frac{1}{2}\ln_{q}n+\frac{1}{2-q}-\frac
{1}{8} & \text{if}\quad q\neq2
\end{array}
\right.  .\label{kekka2}%
\end{align}
From (\ref{kekka1}) and (\ref{kekka2}), we obtain%
\begin{align}
\frac{n}{2-q}\ln_{q}n-\frac{n}{2-q}+\frac{1}{2}\ln_{q}n+\frac{1}{2-q}-\frac
{1}{8}  & <\ln_{q}\left(  n!_{q}\right) \nonumber \\
& <\frac{n}{2-q}\ln_{q}n-\frac{n}{2-q}+\frac{1}{2}\ln_{q}n+\frac{1}%
{2-q}\text{\quad}\left(  \text{if}\quad q\neq2\right)  ,\nonumber \\
& \label{main result1-1}\\
n-\ln n-\frac{1}{2n}-\frac{5}{8}  & <\ln_{q}\left(  n!_{q}\right)  <n-\ln
n-\frac{1}{2n}-\frac{1}{2}\text{\quad}\left(  \text{if}\quad q=2\right)
.\label{main result1-2}%
\end{align}
This result also shows%
\begin{align}
\ln_{q}\left(  n!_{q}\right)   & =\frac{n}{2-q}\ln_{q}n-\frac{n}{2-q}+O\left(
\ln_{q}n\right)  \text{\quad}\left(  \text{if}\quad q\neq2\right)
,\label{main result2-1}\\
\ln_{q}\left(  n!_{q}\right)   & =n-\ln n+O\left(  1\right)  \text{\quad
}\left(  \text{if}\quad q=2\right)  .\label{main result2-2}%
\end{align}

\section{Conclusion}

We present the $q$-Stirling's formula in Tsallis statistics with the proof and
some numerical results. Moreover, some inequalities and equalities concerning
the $q$-Stirling's formula are shown. The present $q$-Stirling's formula can
be applied to many scientific fields in Tsallis statistics as similarly as the
usual Stirling's formula. In particular, the $q$-Stirling's formula plays
important roles in Tsallis statistics when we want to clarify the difference
between the studies $q=1$ and $q\neq1$. Some typical and important
applications of the $q$-Stirling's formula in Tsallis statistics will be
presented in our forthcoming paper \cite{ST04b}, which reveals the surprising
mathematical structure behind Tsallis statistics.

\begin{acknowledgement}
The author would like to thank Professor Yoshinori Uesaka and Professor Makoto
Tsukada for their valuable comments and discussions.
\end{acknowledgement}

\appendix

\section{The proof of (16)}

In this appendix, we prove that $\exp_{q}\left(  x\right)  $ is derived from
$\underset{n\rightarrow \infty}{\lim}\left(  1+\frac{x}{n}\right)
^{\otimes_{q}^{n}}$ using the $q$-product only.

Using the $q$-product (\ref{def of q-product}),\ $\left[  \left(  1+\frac
{x}{n}\right)  ^{\otimes_{q}^{n}}\right]  ^{1-q}$is explicitly expanded as%
\begin{equation}
\left[  \left(  1+\frac{x}{n}\right)  ^{\otimes_{q}^{n}}\right]
^{1-q}=n\left(  1+\frac{x}{n}\right)  ^{1-q}-\left(  n-1\right)
.\label{app-tenkai}%
\end{equation}
The power series expansion for the right side of (\ref{app-tenkai}) about the
point $\frac{x}{n}=0$ is%
\begin{align}
n\left(  1+\frac{x}{n}\right)  ^{1-q}-\left(  n-1\right)   & =n\left[
1+\left(  1-q\right)  \frac{x}{n}-\frac{1}{2}\left(  1-q\right)  q\left(
\frac{x}{n}\right)  ^{2}+O\left(  \left(  \frac{x}{n}\right)  ^{3}\right)
\right]  -\left(  n-1\right) \nonumber \\
& =1+\left(  1-q\right)  x-\frac{1}{2}\left(  1-q\right)  q\frac{x^{2}}%
{n}+O\left(  \frac{x^{3}}{n^{2}}\right)  .
\end{align}
Taking $n\rightarrow \infty$ on the both sides yields%
\begin{equation}
\underset{n\rightarrow \infty}{\lim}\left[  n\left(  1+\frac{x}{n}\right)
^{1-q}-\left(  n-1\right)  \right]  =1+\left(  1-q\right)  x.
\end{equation}
Therefore, if $1+\left(  1-q\right)  x>0$, then
\begin{align}
\underset{n\rightarrow \infty}{\lim}\left(  1+\frac{x}{n}\right)  ^{\otimes
_{q}^{n}}  & =\underset{n\rightarrow \infty}{\lim}\left[  n\left(  1+\frac
{x}{n}\right)  ^{1-q}-\left(  n-1\right)  \right]  ^{\frac{1}{1-q}}\\
& =\left[  1+\left(  1-q\right)  x\right]  ^{\frac{1}{1-q}}\\
& =\exp_{q}\left(  x\right)  .
\end{align}

\end{document}